\documentclass[runningheads]{llncs}
\usepackage[T1]{fontenc}
\usepackage{graphicx}
\usepackage{booktabs}
\usepackage[misc]{ifsym}

\usepackage[hyphens]{url}
\usepackage{amsmath}
\usepackage{amssymb}
\usepackage{tabularx}
\usepackage{ulem}
\usepackage{multirow}
\usepackage{makecell}
\usepackage{listings}
\usepackage{mwe}

\tocauthor{Xiaolu Chen, Jie Bao, Haojie Wu, Zhen Chen, Yong Liao}
\toctitle{Caption Injection for Optimization in Generative Search Engine}

\begin{document}

\title{Caption Injection for Optimization in Generative Search Engine}

\titlerunning{Caption Injection}

\author{Xiaolu Chen\inst{1}\orcidID{0009-0001-9687-9923} \and
Jie Bao\inst{1}\and
Haojie Wu\inst{1}\and
Zhen Chen\inst{1}\Letter\and
Yong Liao\inst{1}}

\authorrunning{X. Chen et al.}


\institute{
University of Science and Technology of China, Hefei, Anhui 230026, China
\email{\{xiaoluchen, baojie1996, hjwuu, cz2016\}@mail.ustc.edu.cn, yliao@ustc.edu.cn}
}

\maketitle              

\begin{abstract}

Generative Search Engine (GSE) leverages the Retrieval-Augmented Generation (RAG) technique and the Large Language Model (LLM) to integrate multi-source information and provide users with accurate and comprehensive responses. Unlike traditional search engines that present results in ranked lists, GSE shifts users’ attention from sequential browsing to content-driven subjective perception, not only driving a paradigm shift in information retrieval but also highlighting the importance of enhancing the subjective visibility of content in generative search. In this context, Generative Search Engine Optimization (G-SEO) methods have emerged as a new research focus. With the rapid advancement of Multimodal Retrieval-Augmented Generation (MRAG) techniques, GSE can now efficiently integrate text, images, audio, and video, producing richer responses that better satisfy complex information needs. Existing G-SEO methods, however, remain limited to text-based optimization and fail to fully exploit multimodal data. To address this gap, we propose Caption Injection, the first multimodal G-SEO approach, which extracts captions from images and injects them into textual content, integrating visual semantics to enhance the subjective visibility in generative search. We systematically evaluate Caption Injection on MRAMG, a benchmark for MRAG, under both unimodal and multimodal settings. Experimental results show that Caption Injection significantly outperforms text-only G-SEO baselines under the G-EVAL metric, effectively improving the subjective visibility of content perceived by users, and demonstrating the practical benefits of multimodal information in G-SEO. The source code for this work is openly available at https://github.com/GrayChan04/Caption-Injection. 

\end{abstract}

\section{Introduction}

\begin{figure}[t]
\includegraphics[width=\textwidth]{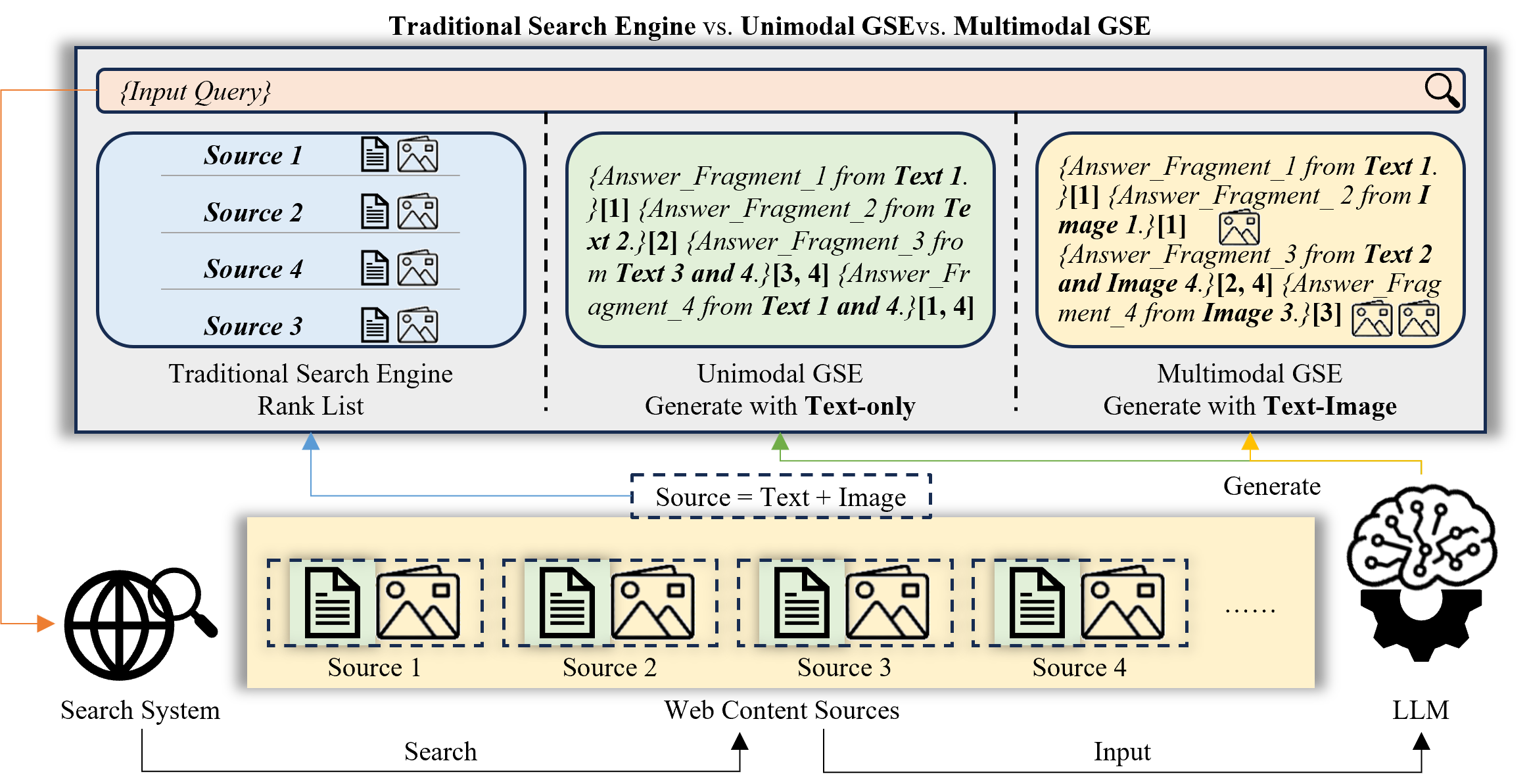}
\caption{Comparison of result presentation across different types of search engines. Traditional search engines (blue section on the left) display retrieved web content sources in a ranked list, where higher-ranked results are typically more relevant to the query. GSE retrieve relevant content sources and leverage LLM to generate comprehensive responses with cited references. Compared with unimodal GSE (green section in the middle) that process only textual information, multimodal GSE (yellow section on the right) jointly interpret textual and visual information, producing responses with richer semantics and higher information density.} \label{fig1}
\end{figure}
Generative Search Engine (GSE) combines Retrieval-Augmented Generation (RAG) techniques with Large Language Model (LLM), transforming how users access information. As shown in Fig.~\ref{fig1}, unlike traditional search engines that present results as ranked lists, GSE can understand user queries, select the most relevant information snippets, and generate integrated responses while providing citation sources. In this generative retrieval paradigm, users can directly perceive information that matches their needs without sequentially browsing search results, thereby improving information acquisition efficiency. As user information needs continue to grow, the development of Multimodal Retrieval-Augmented Generation (MRAG) further enhances GSE’s ability to process and understand both visual and textual information, enabling richer responses and establishing generative search as an increasingly important channel for information access.

Similar to traditional Search Engine Optimization (SEO) in conventional retrieval settings, Generative Search Engine Optimization (G-SEO) aims to increase the user-perceived prominence of content sources in GSE responses, i.e., to improve the subjective visibility of content in generated outputs. However, GSE response is produced by LLM, and its generation depends jointly on retrieval selection and language modeling mechanisms, making content visibility more complex and difficult to control. Although traditional SEO techniques have matured \cite{YALCIN2010Whatissearchengine,ALMUKHTAR2021SEARCHENGINEOPTIMIZATION}, they are primarily designed for keyword-based ranking systems and cannot be directly applied to generative retrieval scenarios. Recently, some studies have explored optimization strategies for GSE \cite{Aggarwal2024geo,chen2025beyondkeywordsdrivinggenerativesearch,pfrommer2024RankingManipulation}. Existing methods mostly focus on text-level optimizations, such as injecting specific text \cite{Kai2023NotWhatYou,kumar2024manipulatinglargelanguagemodels} or restructuring textual semantics \cite{chen2025roleaugmentedintentdrivengenerativesearch,lüttgenau2025BeyondSEO}. However, as GSE increasingly adopts MRAG architectures to support multimodal retrieval, these text-only approaches cannot fully leverage the abundant visual information, limiting their optimization potential in multimodal generative retrieval environments.

To address this limitation, we propose Caption Injection, a G-SEO method for multimodal scenarios. Our key insight is to transform image information into natural language descriptions and inject them into textual content, enabling the fusion of visual and textual semantics. This approach enhances the semantic representation of content and improves its visibility in GSE-generated responses. We conduct a systematic evaluation of Caption Injection under both unimodal and multimodal GSE settings. Experimental results demonstrate that Caption Injection not only maintains stable performance across both settings but also significantly outperforms existing text-only G-SEO strategies in multimodal GSE environments, showing strong cross-modal adaptability.
In summary, our contributions are as follows:
\begin{enumerate}
    \item We extend the G-SEO task from a unimodal to a multimodal setting, systematically defining the multimodal G-SEO problem and constructing a corresponding evaluation framework.
    \item We propose the first multimodal G-SEO method, Caption Injection, which fuses visual and textual semantics via visual information injection, effectively enhancing the subjective visibility of content in GSE responses.
    \item We conduct systematic experiments on the MRAG benchmark dataset MRAMG \cite{Yu2025MRAMGBench}, showing that Caption Injection achieves stable and significant improvements in both retrieval scenarios, validating the effectiveness of multimodal semantic fusion for the G-SEO task.
\end{enumerate}

\section{Related Work}

\subsection{Text-based Generative Search Engine Optimization}
SEO techniques have become relatively mature \cite{Roumeliotis2022AnEffectiveSEO,YALCIN2010Whatissearchengine,ALMUKHTAR2021SEARCHENGINEOPTIMIZATION}, typically influencing search result rankings through multiple factors, including keywords \cite{Kanara2024PythonDrivenKeywordAnalysis,vadlapati2024autotrendykeywords}, webpage structure \cite{Chodak2024LargeLanguageModelsforSearch,Shaffi2022SearchEngineOptimizationbyusing}, and ranking strategies \cite{Bardas2025AutomaticDocumentEditing}. However, when LLM is used as the response generation component in GSE, its probabilistic generation mechanism introduces implicit semantic biases, causing search results to be influenced by the model’s semantic priors. This weakens the effectiveness of traditional SEO methods based on explicit signals, shifting the optimization problem in generative search from explicit signal control to implicit semantic alignment. To address this challenge, G-SEO \cite{Aggarwal2024geo} has been proposed to enhance the visibility of content sources in generative search environments. Existing methods primarily focus on semantic interventions at the text level and can be categorized into two types: content rewriting-based optimization and prompt injection-based optimization. The former, represented by GEO \cite{Aggarwal2024geo}, reconstructs textual semantics to improve alignment with generative mechanisms through approaches such as fine-tuning pretrained models \cite{lüttgenau2025BeyondSEO}, and constructing multi-role intent-aware prompts (Role-Augmented Intent-Driven Generative Search Engine Optimization). The latter guides LLM generation by injecting carefully designed prompts \cite{kumar2024manipulatinglargelanguagemodels,pfrommer2024RankingManipulation,nestaas2024adversarialsearchengineoptimization}, with studies showing that even minor prompt modifications can significantly influence the exposure of content sources in generated responses.

\subsection{Multimodal Content Understanding}
The widespread availability of multimodal data has driven rapid advances in multimodal learning. \cite{Zhu2024Vision+X,Xu2023MultimodalLearningWithTransformers} As a core supporting technology for multimodal GSE, MRAG maps images into the textual semantic space to assist response generation \cite{chen2022murag,Yu2025MRAMGBench,zhu2025murar}. Therefore, multimodal G-SEO tasks can leverage existing research to enable unified semantic space fusion that highlights key information while suppressing noise. Vision-language pretraining models such as CLIP \cite{Radford2021LearningTransferableVisualModels} and the BLIP series \cite{Chaudhuri2022blip,Krause2023blip2} learn cross-modal representations to integrate vision and language, providing knowledge support for downstream tasks \cite{caffagni2024TheRevolutionofMultimodal,zhang2025unifiedmultimodalunderstandinggeneration}. Image captioning, a canonical task \cite{Li2024Pre-TrainedLanguageModelsforTextGeneration,caffagni2024TheRevolutionofMultimodal}, aims to generate fine-grained visual descriptions. Recent studies optimize caption generation via positive-negative sample strategies \cite{xu2024altogether,Rotstein2024FuseCap}, while Capsfusion \cite{Yu2024CapsFusion} and VeCLIP \cite{lai2024VeCLIP} explicitly inject visual knowledge to produce accurate descriptions, providing important inspiration for multimodal G-SEO. Multimodal retrieval and MRAG emphasize cross-modal semantic matching and joint reasoning. Common approaches include mapping images to the textual space and combining them with query text, integrating multiple image captions and context to generate comprehensive text for improved retrieval accuracy \cite{zhu2024EnhancingInteractiveImageRetrieval,zhu2025murar,Yu2025MRAMGBench}, or encoding images and jointly decoding them with text to generate responses \cite{chen2022murag}. These methods demonstrate the critical role of cross-modal semantic alignment and fusion in generative retrieval systems. Moreover, some work about higher-level cross-modal intention understanding further highlight the importance of semantic fusion in complex understanding tasks.  \cite{ye2023Cross-ModalityPyramidAlignment,Yang2025UncertainMultimodalIntention,Sun2024ContextualAugmentedGlobalContrast} 

Despite progress in text-only G-SEO, existing methods are almost entirely confined to textual data, overlooking the rich semantic information present in visual modality, which limits optimization in MRAG scenarios. To address this issue, we propose Caption Injection, a method that injects image semantics into textual content to enable cross-modal optimization, providing a more comprehensive and fine-grained optimization perspective.

\section{Method}
\subsection{Generative Search Engine Optimization Assumption}
To clarify the research context and assumptions, we define the G-SEO task within GSE setting. The GSE workflow is illustrated in Figure 1: given a user query $q$, the system first retrieves a set of content sources $S=Retrieval(q)=\{s_1, s_2, \dots, s_N\}$, and then integrates these sources through an internal LLM to generate a response $r=generate(S,q)$. The response consists of sentences with supporting references $r=\{l_1, l_2, \dots, l_m\}$, where each sentence $l_k(1 \leq k \leq m)$ corresponds to one or more referenced content sources $S_t \subseteq S(1 \leq |S_t| \leq N)$. In the unimodal GSE setting, only textual content sources are used to generate natural language responses. In the multimodal GSE setting, both textual content and image captions are jointly modeled, while some sentences may also include semantically relevant images to enrich content presentation and perceptual understanding. Since GSE responses are presented as coherent passages rather than ranked lists, user attention is largely guided by subjective preference. Accordingly, we introduce subjective visibility to quantify the distribution of user attention and perceptual bias across content sources, which serves as a core metric for G-SEO. Multimodal G-SEO constitutes a new optimization challenge:  LLM must model complex semantic interactions between visual and textual information, which differs fundamentally from purely textual modeling. Building upon prior work \cite{Aggarwal2024geo,chen2025roleaugmentedintentdrivengenerativesearch} in terms of evaluation metrics and GSE scenario design, we extend these settings to multimodal environments, establishing a unified task framework for multimodal G-SEO research.

\subsection{Caption Injection}
\begin{figure}[t]
\centering
\includegraphics[width=\textwidth]{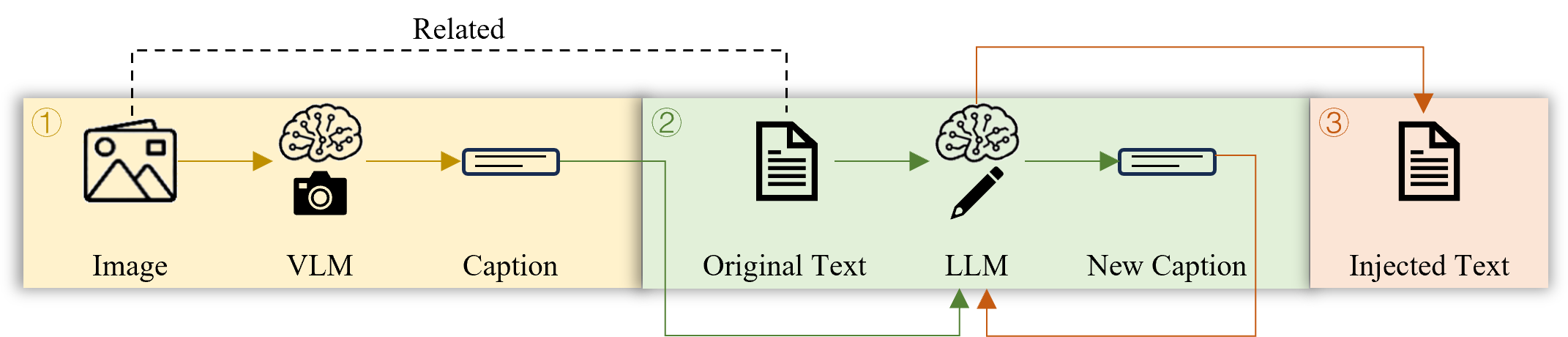}
\caption{Illustration of the Caption Injection pipeline. The image of the web content source is first captioned by the visual-language models, mapping visual representations into the natural language space. The textual content is then injected with the rewritten caption leveraging the LLM, enabling G-SEO with integrated multimodal information.} \label{fig2}
\end{figure}

Consistent with previous G-SEO studies, we treat text content as the primary optimization target, since it dominates response generation in GSE. However, real-world content is often multimodal, and images can provide important information to supplement textual semantics. To extend G-SEO to multimodal scenarios, we propose Caption Injection, which extracts image captions relevant to text semantics and naturally incorporates them into the text, achieving cross-modal semantic enhancement.
Unlike content-rewriting G-SEO methods, image semantics are often local and sparse and cannot independently drive optimization. Therefore, we treat them as a supplementary signal, used only to enhance text segments related to visual content, rather than rewriting the entire text. Inspired by Capsfusion \cite{Yu2024CapsFusion} and VeCLIP \cite{lai2024VeCLIP}, we organize the visual-semantic integration process into three stages: (1) Structural Generation, (2) Alignment Refinement, and (3) Semantic Injection, as illustrated in Fig.~\ref{fig2}.
\subsubsection{Structural Generation}
In real scenarios, image captions are often missing or incomplete, making them insufficient for direct semantic enhancement. We first generate structural captions as stable representations of core visual semantics. Each image is structured using the object–action–scene (O-A-S) triad, which forms the foundational semantic units in visual understanding \cite{bretti2021Zero-shotactionrecognition,Wang2023Contextunderstandingincomputervision}. Given an image $I$, a vision-language model (VLM) extracts its structured semantic representation: 
\begin{equation}
C_s = S(I)=\{o, a, s\}
\end{equation}
where $S$ denotes the structured semantic extraction operation, and $o$, $a$, $s$ correspond to the object, action, and scene semantics in the image.

\subsubsection{Alignment Refinement}
To strengthen the semantic alignment between image descriptions and text, we perform conditional refinement and expansion of the structural caption, using $C_s$ as a semantic framework and incorporating the text context $T$:
\begin{equation}
C_r = R(T, C_s).
\end{equation}
Here, $R$ denotes the refinement operation. During this process, we extract key knowledge from the text that is most relevant to the core elements of the structural caption for semantic expansion, while preserving its original syntactic structure as a semantic anchor. This guides the LLM with visual semantic cues and reduces potential semantic drift. The resulting refined caption $C_r$ provides more complete semantic coverage and is naturally aligned with the text context.

\subsubsection{Semantic Injection}
Finally, the refined caption $C_r$ is integrated into the text $T$ to enhance multimodal semantics. The LLM automatically determines the optimal insertion points based on contextual semantic dependencies, seamlessly embedding the image description into the text:
\begin{equation}
T' =J(T, C_r)
\end{equation}
where $T'$ represents the text after integrating visual semantics, and $J$ denotes the context-aware semantic injection operation. This process preserves the original text structure while explicitly transferring visual information, enhancing both semantic completeness and informational richness.

Through the three-stage Generate–Refine–Inject pipeline, Caption Injection achieves controlled visual-semantic integration while maintaining text-dominated optimization. The entire process is implemented via prompt engineering, with relevant prompt templates provided in Appendix A.

\section{Experiments and Results}
\subsection{Experimental Setting}
\subsubsection{Generative Search Engine Simulation}
To fairly evaluate the performance of G-SEO methods, we designed a unified simulation framework covering both unimodal and multimodal scenarios, and simplified the GSE workflow to single-turn response generation to exclude retrieval-stage interference. Notably, image captions are not used as input in this setting. In the unimodal scenario, responses are generated using only the content source text and LLM, simulating a typical text-driven generative process. In the multimodal scenario, we compared several MRAG implementations \cite{Yu2025MRAMGBench,chen2022murag,zhu2025murar} in terms of retrieval accuracy, recall, and the relevance and fluency of generated responses. We found that the combination of ``content source text + image caption + LLM'' is more stable, and therefore adopted it for the simulation. To reduce the impact of model hallucinations on experimental results, we employed the low-hallucination open-source model GLM-4-9B\footnote{https://github.com/vectara/hallucination-leaderboard/} for response generation, providing stable outputs for an objective assessment of optimization effects. The prompts used for generation are detailed in Appendix B. 

\subsubsection{Datasets}
Multimodal G-SEO is a novel research direction, and no dedicated public dataset currently exists. Since the RAG task serves as the upstream stage for GSE generation, its query–content pair structure closely aligns with generative retrieval scenarios. Therefore, we adopt the MRAG benchmark dataset MRAMG \cite{Yu2025MRAMGBench}, which is more representative in terms of modality balance and cross-domain coverage, for our experiments. This dataset provides both text and image content sources across multiple domains, supporting diverse inputs and complex content requirements. MRAMG consists of 4,800 query-content pairs drawn from six distinct domains, categorized into three difficulty levels according to task complexity: 1) Easy-level Web data: Includes the Wit, Wiki, and Web datasets derived from Wikipedia articles, totaling 1,850 samples. These datasets focus on fundamental text–image understanding within relatively simple and homogeneous contexts. 2) Medium-level academic data: includes the Arxiv dataset containing 200 samples collected from papers published on arXiv between 2023 and 2024. Each sample often includes multiple figures, designed to evaluate cross-modal semantic fusion and academic knowledge generation. 3) Hard-level lifestyle data: includes Recipe and Manual datasets with 2,750 samples. Furthermore, while the average text length of the other datasets is generally on the order of hundreds of characters, the Manual dataset contains texts averaging thousands of characters, introducing additional comprehension challenges. During preprocessing, we strictly follow the original data structure and splits, preserving all annotations, modality ratios, and image resolutions, without any additional cleaning, relabeling, or augmentation, to ensure the reproducibility of experimental results.

\subsubsection{Baselines}
To comprehensively evaluate the effectiveness of Caption Injection in enhancing content-source visibility in both unimodal and multimodal scenarios, we categorized existing text-only G-SEO strategies \cite{Aggarwal2024geo} and selected representative methods as baselines: (1) Traditional SEO, which simulates keyword placement optimization in conventional SEO; (2) Fluency Expression Optimization, which refines and rewrites text to improve fluency without altering the core semantics; and (3) Statistics- and Quotation-based Addition Optimization, which selectively incorporate statistical data or famous quotations to enrich the text and enhance its persuasiveness.
All baselines and Caption Injection are implemented using GLM-4-9B. In the implementation of Caption Injection, we employ Qwen-2.5-VL-7B\footnote{https://github.com/BradyFU/Awesome-Multimodal-Large-Language-Models/tree/Evaluation?tab=readme-ov-file} to automatically generate captions for images lacking descriptions. The generated results are then randomly sampled and independently evaluated by multiple domain experts for semantic consistency with the corresponding images to ensure quality. All prompts and LLM parameters used in the experiments are provided in the open-source code\footnote{https://github.com/GrayChan04/Caption-Injection}.

\subsubsection{Evaluation Metrics}
In the G-SEO task, users typically access content sources indirectly through generated responses. Accordingly, the degree to which a content source is represented in generated outputs serves as an important proxy for evaluating the effectiveness of G-SEO methods, measuring its subjective visibility and characterizing its prominence and influence within generated results. We quantify it using the G-SEO–adapted G-EVAL 2.0 \cite{Aggarwal2024geo,chen2025roleaugmentedintentdrivengenerativesearch}. G-EVAL 2.0 evaluates content sources across seven dimensions: relevance, influence, diversity, uniqueness, click-follow probability, positional salience, and content volume. Each dimension is scored on a 0–5 scale, and the overall subjective visibility is computed as the average across the seven dimensions. The optimization effect of a G-SEO method is measured by relative improvement, calculated as:
\begin{equation}
\text{improvement}_{(s, s')} =
\frac{\text{impression}_{s'}(r') - \text{impression}_{s}(r)}
{\text{impression}_{s}(r) + 1}
\times 100\%
\end{equation}
where $s$ and $r$ denote the content source and GSE response before optimization, $s'$ and $r'$ denote the counterparts after optimization, and $impression_s(r)$ represents the score of a sub-dimension or the average subjective visibility. To reduce randomness, each sample’s improvement is averaged over three independent experimental runs to determine the final performance. Although G-EVAL 2.0 is not a human-annotated gold standard, it provides a structured and reproducible framework for assessing subjective visibility. The relative improvement should therefore be interpreted as a comparative measure across G-SEO methods, rather than an absolute estimate of user-perceived visibility. 

\begin{table}[t]
\centering
\caption{Average relative improvement (\%) in subjective visibility achieved by different G-SEO methods on the MRAMG dataset under both unimodal and multimodal generative retrieval settings. The reported values indicate the average improvement compared to the original content sources, and the average score across datasets is used as the final performance metric for each optimization method. The best results are highlighted in bold, while the second-best results are underlined.}
\label{tab1}
\begin{tabularx}{\textwidth}{l *{7}{>{\centering\arraybackslash}X}}
\toprule
\multicolumn{8}{c}{\textbf{Unimodal}} \\
\midrule
Method & Arxiv & Manual & Recipe & Web & Wiki & Wit & Average \\
\midrule
Tran\_SEO   & -0.74 & \uline{-8.93} &  0.64 & -3.05 & -0.14 & -1.58 & -2.30 \\
Flue\_Expr  &  \uline{0.98} & \textbf{-5.88} &  \textbf{2.21} &  \uline{0.04} &  0.36 &  0.09 & \textbf{-0.37} \\
Quat\_Addi  &  0.55 & -11.38 & 1.97 &  \textbf{0.46} &  \textbf{0.97} &  \textbf{0.70} & -1.12 \\
Stat\_Addi  & -1.68 & -11.26 & -1.75 & -4.50 & -1.93 & -3.63 & -4.13 \\
Capt\_Inje  &  \textbf{1.39} & -10.51 & \uline{2.10} & -0.36 &  \uline{0.70} &  \uline{0.61} & \uline{-1.01} \\
\midrule
\multicolumn{8}{c}{\textbf{Multimodal}} \\
\midrule
Method & Arxiv & Manual & Recipe & Web & Wiki & Wit & Average \\
\midrule
Tran\_SEO   & -0.90 & \uline{-0.49} &  0.64 & -1.96 & -0.57 & -0.65 & -0.66 \\
Flue\_Expr  &  0.48 &  \textbf{1.54} &  1.85 & -0.07 &  0.58 & -0.15 &  \uline{0.71} \\
Quat\_Addi  &  \uline{0.78} & -5.74 &  \textbf{2.21} &  \uline{0.61} &  \uline{1.02} &  \uline{0.65} & -0.08 \\
Stat\_Addi  & -1.60 & -3.54 & -0.58 & -3.10 & -1.77 & -2.15 & -2.12 \\
Capt\_Inje  &  \textbf{1.44} & -0.81 &  \uline{1.90} &  \textbf{1.33} &  \textbf{1.24} &  \textbf{1.60} &  \textbf{1.12} \\
\bottomrule
\end{tabularx}
\end{table}
\subsection{Results and Analysis}
\subsubsection{Main Result}
We systematically evaluated Caption Injection against several text-only G-SEO methods on the MRAMG benchmark dataset for the subjective visibility optimization task, comparing performance under both unimodal and multimodal GSE scenarios. Complete experimental results are provided in Appendix C. Tab.~\ref{tab1} reports the average relative improvement in subjective visibility across all datasets. Overall, G-SEO methods exhibit more stable optimization performance in the unimodal scenario than in the multimodal scenario, indicating that the introduction of multimodal information increases the difficulty of G-SEO tasks, as its added semantic complexity may reduce the effectiveness of text-only strategies.
In the unimodal scenario, most G-SEO methods achieve relatively stable improvements. Fluency Expression Optimization attains the best performance, with a relative improvement of $-0.37\% (\pm 0.03\%)$, while Caption Injection ranks second at $-1.01\% (\pm 0.06\%)$. These results suggest that, when relying solely on textual information, enhancing textual fluency can effectively improve the model’s comprehension of content. We further examined the multimodal scenario, where Caption Injection significantly outperforms all text-only baselines, achieving a relative improvement of $1.12\% (\pm 0.02\%)$, which is $0.41\% (\pm 0.03\%)$ higher than the next-best method, Fluency Expression Optimization, with statistical tests confirming the robustness of this improvement. This indicates that incorporating descriptions aligned with visual content provides additional semantics, thereby enhancing optimization effectiveness and highlighting the potential value of multimodal information in G-SEO tasks.
At the dataset level, all methods show performance degradation on MRAMG-Manual, with negative relative improvements in both scenarios. This decline is likely due to the dataset’s extremely long texts, which have an average length of 6,365.4 characters \cite{Yu2025MRAMGBench}. Excessively long texts may dilute the density of key information, increasing the difficulty of identifying and optimizing core semantics. Nevertheless, Fluency Expression Optimization still performs relatively well on this dataset, likely because fluency enhancements improve the readability of complex technical texts, helping the GSE better interpret highly specialized content. This further illustrates that dataset characteristics, such as text length and domain specificity, significantly affect the optimization performance of G-SEO methods.
Moreover, we observe some discrepancies with previous studies \cite{Aggarwal2024geo,chen2025roleaugmentedintentdrivengenerativesearch}. For instance, Statistics-based Addition Optimization consistently underperforms across datasets. This suggests that the effectiveness of optimization strategies partially depends on their alignment with the domain-specific content distribution.
Overall, Caption Injection demonstrates stable performance in subjective visibility optimization and shows a clear advantage in multimodal scenarios. These findings indicate that multimodal information can serve as an important signal for G-SEO and emphasize the importance of further exploring multimodal-aware G-SEO methods. A detailed discussion of limitations and potential future work is provided in Appendix D.

\begin{figure}[t]
\centering
\includegraphics[width=\textwidth]{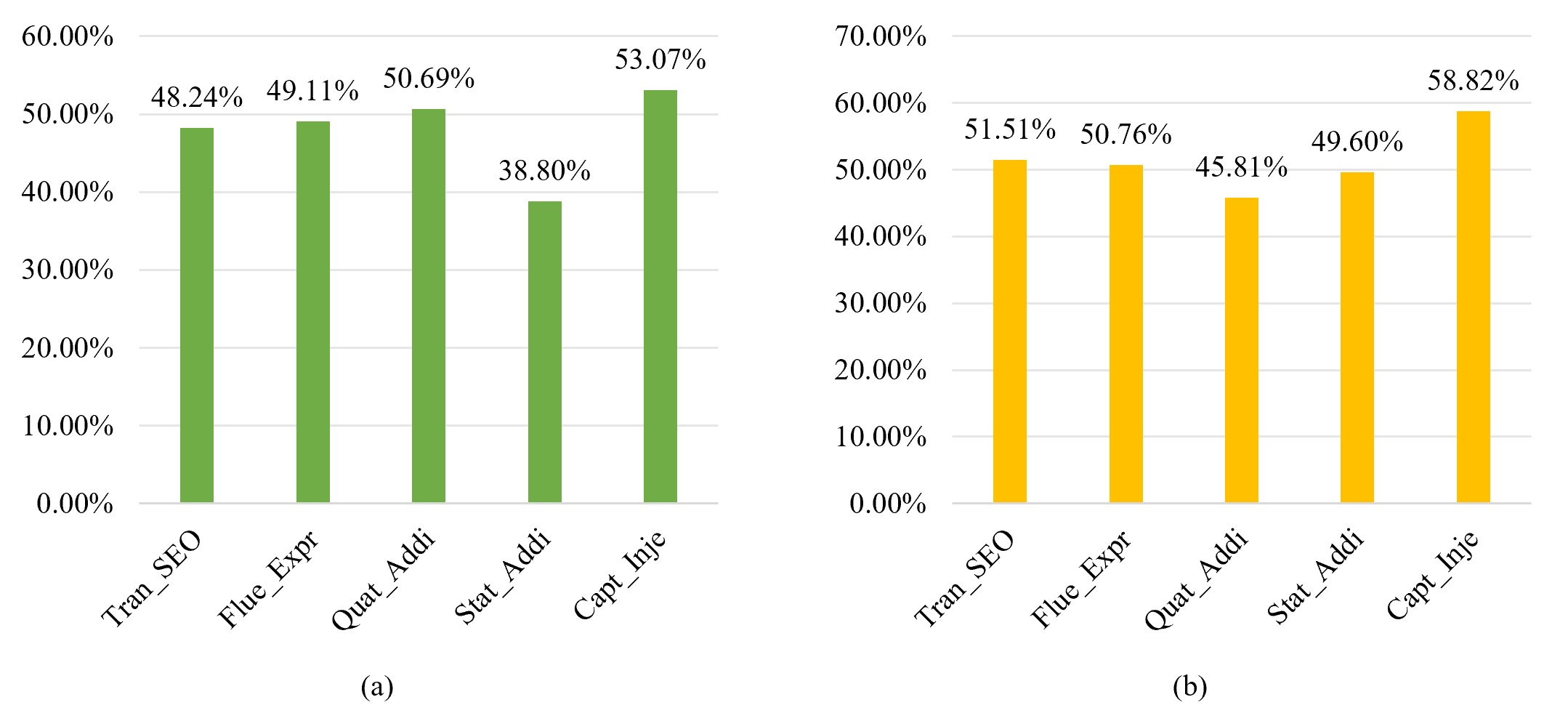}
\caption{Adaptability analysis of different G-SEO methods on the MRAMG dataset under (a) unimodal and (b) multimodal generative retrieval settings.} \label{fig3}
\end{figure}

\begin{figure}[t]
\centering
\includegraphics[width=0.8\textwidth]{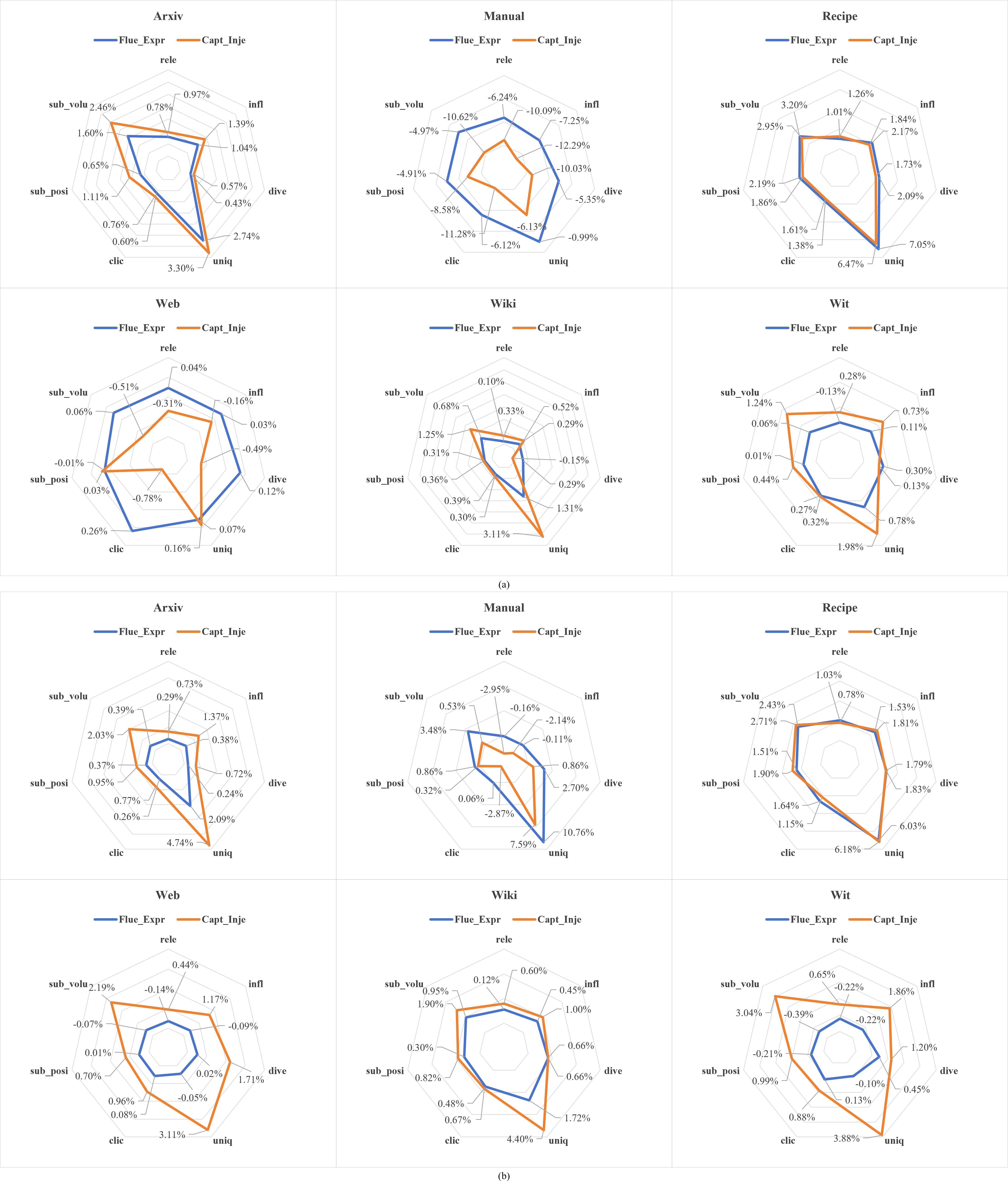}
\caption{Comparison of the contributions of Fluency Expression Optimization and Caption Injection to relative improvements (\%)  in the subjective visibility of content sources under (a) unimodal and (b) multimodal setting. } \label{fig4}
\end{figure}
\subsubsection{Adaptability Analysis}
We considered an effective optimization event whenever the average relative subjective visibility improvement was positive on MRAMG datasets \cite{Yu2025MRAMGBench}. The average effective optimization rate across all datasets was used as an adaptability metric to quantify the generalization capability. The analysis results are shown in Fig.~\ref{fig3}.
Overall, Caption Injection demonstrated the most robust and comprehensive adaptability. Its effective optimization rate in the multimodal setting increased by $5.75\%$ compared to the unimodal setting, highlighting its broad applicability in multimodal G-SEO tasks.
Notably, rankings by effective optimization rate did not fully align with rankings by subjective visibility improvement. In other words, larger improvements do not necessarily exhibit stronger adaptability. For example, Fluency Expression Optimization had slightly lower effective optimization rates than Quotation-based Addition Optimization and Traditional SEO under both modality settings, yet achieved higher relative improvement. This indicates that adaptability (generalization capability) and optimization strength (improvement magnitude) are two independent yet important performance dimensions for G-SEO methods.
Additionally, Statistics-based Addition Optimization and Quotation-based Addition Optimization showed the largest changes between unimodal and multimodal settings, at $4.88\%$ and $10.8\%$ respectively, reflecting the varying difficulty across G-SEO tasks. While effective optimization rates are higher in the multimodal setting, relative improvements are generally lower, indicating the increased challenge.
In summary, G-SEO methods require balancing adaptability and optimization strength to address diverse data scenarios. The comprehensive adaptability of Caption Injection further confirms its potential in multimodal G-SEO .

\subsubsection{Contribution Analysis}
To validate the role of multimodal information in the G-SEO task, we select Fluency Expression Optimization as the representative text-only baseline and compare it with Caption Injection in terms of their contributions to relative improvements across different subjective visibility dimensions. As shown in Fig.~\ref{fig4}, Caption Injection achieves significantly higher contributions to the uniqueness dimension compared to the relatively balanced distribution of Fluency Expression Optimization in both settings. This effect is particularly pronounced on the MRAMG-Manual dataset: in the unimodal setting, uniqueness exceeds the imension, influence, by $6.16\%$, and in the multimodal setting, it exceeds relevance by $10.54\%$, indicating that multimodal information effectively supplements missing textual content. Tab.~\ref{tab2} further reports the changes in contributions across dimensions when migrating from unimodal to multimodal settings. The average incremental improvement for Caption Injection is $2.13\%$, with uniqueness contributing $3.50\%$; in contrast, Fluency Expression Optimization shows an incremental improvement of only $1.07\%$, with uniqueness contributing $1.58\%$. Across almost all sub-dimensions, the incremental gains of Caption Injection surpass those of the baseline. These results indicate that multimodal information enhances the differentiated expression of content sources through additional visual-semantic cues, improving overall subjective visibility and boosting content source uniqueness.
\begin{table}[t]
\centering
\caption{Changes in relative improvements (\%) of Fluency Expression Optimization and Caption Injection across subjective visibility sub-dimensions when transitioning from unimodal to multimodal settings. The method implementations are kept unchanged, with only the task modality adjusted. For each method and dataset, the sub-dimension with the largest incremental improvement is highlighted in bold.}
\label{tab2}
\begin{tabularx}{\textwidth}{l l *{8}{>{\centering\arraybackslash}X}}
\toprule
Dataset & Method & Rele. & Infl. & Dive. & Uniq. & Clic. & \makecell{Sub.\\Posi.} & \makecell{Sub.\\Volu.} & Aver. \\
\midrule
\multirow{2}{*}{Arxiv}
& Flue\_Expr & -0.49 & -0.66 & \textbf{-0.19} & -0.65 & -0.34 & -0.28 & -1.21 & -0.50 \\
& Capt\_Inje & -0.24 & -0.02 & 0.15 & \textbf{1.44} & 0.01 & -0.16 & -0.43 & 0.05 \\
\midrule
\multirow{2}{*}{Manual}
& Flue\_Expr & 6.08 & 7.14 & 8.05 & \textbf{11.75} & 6.18 & 5.77 & 8.45 & 7.42 \\
& Capt\_Inje & 7.14 & 10.15 & 10.89 & \textbf{13.72} & 8.41 & 8.90 & 11.15 & 9.70 \\
\midrule
\multirow{2}{*}{Recipe}
& Flue\_Expr & 0.02 & -0.64 & -0.26 & -1.02 & \textbf{0.03} & -0.68 & -0.77 & -0.36 \\
& Capt\_Inje & -0.48 & -0.03 & \textbf{0.06} & -0.29 & -0.23 & 0.04 & -0.24 & -0.20 \\
\midrule
\multirow{2}{*}{Web}
& Flue\_Expr & -0.18 & -0.12 & -0.10 & -0.12 & -0.18 & \textbf{0.02} & -0.13 & -0.11 \\
& Capt\_Inje & 0.75 & 1.33 & 2.20 & \textbf{2.95} & 1.74 & 0.67 & 2.70 & 1.69 \\
\midrule
\multirow{2}{*}{Wiki}
& Flue\_Expr & 0.02 & 0.16 & 0.37 & \textbf{0.41} & 0.18 & -0.01 & 0.27 & 0.22 \\
& Capt\_Inje & 0.27 & 0.48 & 0.81 & \textbf{1.29} & 0.28 & 0.46 & 0.65 & 0.54 \\
\midrule
\multirow{2}{*}{Wit}
& Flue\_Expr & -0.09 & -0.33 & \textbf{0.15} & -0.88 & -0.14 & -0.22 & -0.45 & -0.24 \\
& Capt\_Inje & 0.37 & 1.13 & 1.07 & \textbf{1.90} & 0.56 & 0.55 & 1.80 & 0.99 \\
\midrule
\multirow{2}{*}{Average}
& Flue\_Expr & 0.89 & 0.93 & 1.34 & \textbf{1.58} & 0.96 & 0.77 & 1.03 & 1.07 \\
& Capt\_Inje & 1.30 & 2.17 & 2.53 & \textbf{3.50} & 1.80 & 1.74 & 2.61 & 2.13 \\
\bottomrule
\end{tabularx}
\end{table}

\subsubsection{Ablation Study}
To evaluate the contribution of refined captions to subjective content visibility, we used structural captions as a baseline, keeping the injection procedure unchanged. For each dataset, we performed single-shot injections under both unimodal and multimodal settings, and measured their performance using the average relative improvement across datasets. The results are shown in Tab.~\ref{tab3}. Refined captions outperform structural captions in both settings, confirming the effectiveness of the Alignment Refinement, likely because fine-grained rewriting provides additional semantics more focused on key segments, facilitating the LLM in capturing relevant information. The advantage of refined captions is more pronounced in the multimodal setting. This is likely due to the interaction between supplementary semantics and input captions, which enhances the LLM’s ability to process multimodal information, improving optimization performance. Tab.~\ref{tab3} also compares single-shot and multi-shot injections. In the unimodal setting, multiple injections significantly improve performance, whereas in the multimodal setting, repeated injections have limited effect. This difference may arise from the LLM’s preference for text versus image captions: in the purely textual scenario, repeated injections reinforce the understanding of shallow concepts; in the multimodal scenario, they increase the attention weight on visual semantics, reducing focus on textual details and thereby limiting optimization. Overall, the study validates refined captions and highlights differences in unimodal and multimodal G-SEO processing, offering insights for multimodal strategy design.
\begin{table}[t]
\centering
\caption{Ablation study results of Caption Injection on MRAMG. The average performance across all datasets is adopted as the performance metric for comparison, illustrating the contributions of the proposed Alignment Refinement and Semantic Injection stages.}
\label{tab3}
\begin{tabularx}{\textwidth}{l *{8}{>{\centering\arraybackslash}X}}
\toprule
\multicolumn{9}{c}{\textbf{Unimodal}} \\
\midrule
Caption Type & Rele. & Infl. & Dive. & Uniq. & Clic. & \makecell{Sub.\\Posi.} & \makecell{Sub.\\Volu.} & Aver. \\
\midrule
\makecell{Structural Caption \\ (single-shot)}
& -1.37 & -1.53 & -1.30 & 1.15 & -1.52 & -1.00 & -0.81 & -1.15 \\
\makecell{Refined Caption \\ (single-shot)}
& -1.26 & -1.33 & -1.37 & 1.48 & -1.54 & -0.80 & -0.54 & -1.01 \\
\makecell{Refined Caption \\ (multi-shot)} 
& -0.88 & -0.93 & -0.79 & 1.52 & -0.97 & -0.49 & -0.27 & -0.63 \\
\midrule
\multicolumn{9}{c}{\textbf{Multimodal}} \\
\midrule
Caption Type & Rele. & Infl. & Dive. & Uniq. & Clic. & \makecell{Sub.\\Posi.} & \makecell{Sub.\\Volu.} & Aver. \\
\midrule
\makecell{Structural Caption \\ (single-shot)}
& -0.30 & 0.41 & 0.56 & 4.52 & -0.21 & 0.67 & 1.79 & 0.71 \\
\makecell{Refined Caption \\ (single-shot)}
& 0.04 & 0.85 & 1.16 & 4.98 & 0.26 & 0.95 & 2.07 & 1.12 \\
\makecell{Refined Caption \\ (multi-shot)} 
& -0.31 & -0.31 & 0.16 & 2.04 & -0.28 & 0.00 & 0.27 & 0.01 \\
\bottomrule
\end{tabularx}
\end{table}

\section{Conclusion}
GSE has shifted users’ information acquisition toward the subjective perception of paragraph-level content, motivating research on G-SEO aimed at enhancing the subjective visibility of content. The advancement of MRAG equips the GSE with multimodal capabilities while posing new G-SEO challenges. To address this, we extend the G-SEO task to multimodal settings and propose Caption Injection, the first multimodal G-SEO method, which transforms image information into supplementary semantics for multimodal optimization. Experiments show that Caption Injection significantly improves the subjective visibility of content sources in both unimodal and multimodal GSE settings, demonstrating the effectiveness of multimodal semantic fusion. Although deep interactions of multimodal features remain unexplored, this work provides a practical approach and supporting evidence for systematically integrating visual information into G-SEO. Future work will focus on fine-grained visual-semantic interactions and LLM preference analysis for robust cross-modal G-SEO optimization strategies.

\section{Limitations}
Although the experimental results support the effectiveness of Caption Injection, several limitations remain. The evaluation setup was informed by existing MRAG studies. Since prior work has consistently shown that LLM-centered architectures achieve strong performance on public benchmarks, this work adopted the same configuration for experiments. However, a systematic comparison against VLMs capable of processing images directly has not been conducted, so the applicability and generalization of Caption Injection across different multimodal generation architectures remain to be validated.
While Caption Injection outperforms other text-based baselines, its average relative improvement remains modest (approximately 1.1\%). In addition, most G-SEO methods achieve substantially lower improvements in multimodal GSE settings compared to unimodal scenarios, highlighting the challenges of multimodal G-SEO. Furthermore, it is unclear whether such gains translate into perceptible user benefits in real-world GSE applications.
Experiments confirm the potential of multimodal information fusion for G-SEO tasks, yet deep interactions between image and text features remain largely unexplored. Currently, the visual-text semantic fusion in Caption Injection is still shallow, limiting the effective use of image information to supplement textual content. This suggests that future work should investigate deeper cross-modal feature integration and construct a unified semantic space.
We further observe that the GSE’s preference for input data may affect the alignment between optimization results and its internal semantic representations, thereby limiting the practical influence of optimization methods on GSE semantic understanding. These findings motivate future research on cross-model optimization strategies, advancing understanding of the GSE black-box mechanism and achieving more effective G-SEO strategies.

\section{Ethical Statement}
We strictly followed the ethical standards of academic research in artificial intelligence, ensuring transparency in data usage, model utilization, and result presentation. All LLMs and datasets used in this work are sourced from open communities and are used in accordance with their licensing terms.
This research is conducted solely to explore and validate methods for optimizing the visibility of web content sources in GSE. The proposed approach is not intended to manipulate search results or to be applied to real-world operations or potential misuse. All outputs generated by the LLM are used exclusively for academic research, without any commercial perposes. 
During the preparation of this work, LLM-based tools were employed solely to improve the clarity and readability of the text and were not used to generate research content or analytical conclusions.


\begin{credits}
\subsubsection{\ackname} This work was supported by the National Science and Technology Innovation 2030 Major Project (Grant No. 2025ZD1502104), and the Anhui Province Science and Technology Key Project (Grant No. 202423l10050033).

\subsubsection{\discintname}
The authors have no competing interests to declare that are relevant to the content of this article. 
\end{credits}

\appendix 

\section*{Appendix}

\section{Caption Injection Prompt}

For clarity, we present only the core prompt logic of Caption Injection process in this section, while the unified system and role configurations used as LLM parameters remain consistent and are not elaborated. Complete prompt templates will be released with our open-source code: https://github.com/GrayChan04/Caption-Injection.

\subsection{Structual Caption Generation Prompt}
\begin{lstlisting}[breakindent=0pt, breakatwhitespace=true]
Generate a concise and objective caption for this image, describing the main objects, actions, and scene present. Do not include any subjective emotions, opinions, or speculation. Output only the caption without any explanation or analysis. Output plain text only, without quotes or parentheses.
\end{lstlisting}

\subsection{Alignment Refinement Prompt}
\begin{lstlisting}[breakindent=0pt, breakatwhitespace=true]
1. Carefully read the source and the original caption, and identify the relevant details.
2. Based on the identified details, replace the core elements of the original caption (who/what/when/where/how):
- If any element is missing in the source, keep the corresponding element from the original caption.
- If there is a conflict, the source takes precedence.
3. Extract the key information from the source that is most relevant to the revised caption, and expand the caption while maintaining logical coherence, ignoring unrelated details.
4. Preserve a sentence structure similar to the original caption, while making appropriate adjustments to enhance expressiveness and appeal. The length of the rewritten caption should be 50%-150% of the original caption (in number of words).
\end{lstlisting}

\subsection{Semantic Injection Prompt}
\begin{lstlisting}[breakindent=0pt, breakatwhitespace=true]
1. Identify the most suitable position in the source for inserting the text naturally.
2. Insert the text only at the position, ensuring smooth and coherent context.
3. Do not delete or modify any other part of the given source or the text, and do not add anything other than the given text.
\end{lstlisting}

\section{Generative Search Engine Response Generation Simulation Prompt}
To simulate the response generation process of the GSE, we design a unified prompt with explicit role definitions and task instructions to ensure consistency in the generated responses. In the unimodal setting, only the query and the content source text are provided as input. In the multimodal setting, a ``text content + image captions + LLM'' input configuration is adopted. To ensure a fair comparison, the multimodal prompt minimally extends the unimodal prompt by incorporating only the necessary multimodal information.
\begin{lstlisting}[breakindent=0pt, breakatwhitespace=true]
You are a comprehensive and objective query answerer.

Using only the given web source contents (*@\underline{and related image captions}@*), write a correct, clear, and high-quality answer to the given query. Your answer must follow these requirements:
- Only include relevant content to the query; do not include any unrelated information
- Be informative, logically rigorous and clear, and also engaging
- Maintain a fair and objective tone; do not express any stance or bias
- Use the same language as the query
- Each sentence in the answer must be followed by a citation in the format [index], where index corresponds to the number of the cited source (*@\underline{or caption}@*) in the given Source Contents (*@\underline{and Image Captions}@*)
- If a sentence cites multiple sources, list multiple separate citations like [1][2] instead of a combined form like [1,2]
\end{lstlisting}
\begin{table}[t]
\centering
\caption{Relative improvement (\%) on subjective visibility sub-dimensions of Caption Injection and text-only baselines on the MRAMG-Arxiv dataset.}
\label{tab4}
\begin{tabularx}{\textwidth}{l *{8}{>{\centering\arraybackslash}X}}
\toprule
\multicolumn{9}{c}{\textbf{Unimodal}} \\
\midrule
Method & Rele. & Infl. & Dive. & Uniq. & Clic. & \makecell{Sub.\\Posi.} & \makecell{Sub.\\Volu.} & Aver. \\
\midrule
Tran\_SEO  & -0.48 & -0.45 & 0.21 & -2.83 & -0.16 & -0.60 & -0.81 & -0.74 \\
Flue\_Expr &  0.78 &  1.04 & 0.43 &  2.74 &  0.60 &  0.65 &  1.60 &  0.98 \\
Quat\_Addi &  0.19 &  0.69 & 0.27 &  1.31 &  0.25 &  0.57 &  1.17 &  0.55 \\
Stat\_Addi & -0.78 & -1.09 & -1.02 & -5.31 & -0.37 & -1.16 & -2.26 & -1.68 \\
Capt\_Inje &  0.97 &  1.39 & 0.57 &  3.30 &  0.76 &  1.11 &  2.46 &  1.39 \\
\midrule
\multicolumn{9}{c}{\textbf{Multimodal}} \\
\midrule
Method & Rele. & Infl. & Dive. & Uniq. & Clic. & \makecell{Sub.\\Posi.} & \makecell{Sub.\\Volu.} & Aver. \\
\midrule
Tran\_SEO  & -0.77 & -0.87 & 0.07 & -2.22 & -0.59 & -0.55 & -1.20 & -0.90 \\
Flue\_Expr &  0.29 &  0.38 & 0.24 &  2.09 &  0.26 &  0.37 &  0.39 &  0.48 \\
Quat\_Addi &  0.46 &  0.67 & 0.73 &  2.03 &  0.71 &  0.66 &  0.85 &  0.78 \\
Stat\_Addi & -1.05 & -1.51 & -0.73 & -3.49 & -0.57 & -1.27 & -2.61 & -1.60 \\
Capt\_Inje &  0.73 &  1.37 & 0.72 &  4.74 &  0.77 &  0.95 &  2.03 &  1.44 \\
\bottomrule
\end{tabularx}
\end{table}
\begin{table}[t]
\centering
\caption{Relative improvement (\%) on subjective visibility sub-dimensions of Caption Injection and text-only baselines on the MRAMG-Manual dataset.}
\label{tab5}
\begin{tabularx}{\textwidth}{l *{8}{>{\centering\arraybackslash}X}}
\toprule
\multicolumn{9}{c}{\textbf{Unimodal}} \\
\midrule
Method & Rele. & Infl. & Dive. & Uniq. & Clic. & \makecell{Sub.\\Posi.} & \makecell{Sub.\\Volu.} & Aver. \\
\midrule
Tran\_SEO  & -8.45 & -10.08 & -6.59 & -8.53 & -8.05 & -7.37 & -9.94 & -8.93 \\
Flue\_Expr & -6.24 &  -7.25 & -5.35 & -0.99 & -6.12 & -4.91 & -4.97 & -5.88 \\
Quat\_Addi & -11.18 & -12.97 & -10.75 & -6.79 & -12.46 & -9.16 & -11.64 & -11.38 \\
Stat\_Addi & -10.18 & -12.26 & -10.10 & -10.96 & -9.17 & -9.47 & -13.55 & -11.26 \\
Capt\_Inje & -10.09 & -12.29 & -10.03 & -6.13 & -11.28 & -8.58 & -10.62 & -10.51 \\
\midrule
\multicolumn{9}{c}{\textbf{Multimodal}} \\
\midrule
Method & Rele. & Infl. & Dive. & Uniq. & Clic. & \makecell{Sub.\\Posi.} & \makecell{Sub.\\Volu.} & Aver. \\
\midrule
Tran\_SEO  & -1.77 & -1.18 &  2.41 &  3.27 & -0.41 & -0.89 & -0.10 & -0.49 \\
Flue\_Expr & -0.16 & -0.11 &  2.70 & 10.76 &  0.06 &  0.86 &  3.48 &  1.54 \\
Quat\_Addi & -6.47 & -7.39 & -4.78 &  1.13 & -7.07 & -4.94 & -5.27 & -5.74 \\
Stat\_Addi & -3.79 & -4.47 & -1.23 & -0.82 & -2.63 & -3.33 & -4.52 & -3.54 \\
Capt\_Inje & -2.95 & -2.14 &  0.86 &  7.59 & -2.87 &  0.32 &  0.53 & -0.81 \\
\bottomrule
\end{tabularx}
\end{table}
\begin{table}[t]
\centering
\caption{Relative improvement (\%) on subjective visibility sub-dimensions of Caption Injection and text-only baselines on the MRAMG-Recipe dataset.}
\label{tab6}
\begin{tabularx}{\textwidth}{l *{8}{>{\centering\arraybackslash}X}}
\toprule
\multicolumn{9}{c}{\textbf{Unimodal}} \\
\midrule
Method & Rele. & Infl. & Dive. & Uniq. & Clic. & \makecell{Sub.\\Posi.} & \makecell{Sub.\\Volu.} & Aver. \\
\midrule
Tran\_SEO  & -0.05 & 0.54 & 1.91 & 2.41 & 0.72 & 0.69 & 0.90 & 0.64 \\
Flue\_Expr & 1.01 & 2.17 & 2.09 & 7.05 & 1.61 & 2.19 & 3.20 & 2.21 \\
Quat\_Addi & 0.90 & 1.67 & 2.21 & 6.31 & 1.46 & 1.70 & 2.51 & 1.97 \\
Stat\_Addi & -0.95 & -1.67 & 0.48 & -3.43 & 0.58 & -1.66 & -3.87 & -1.75 \\
Capt\_Inje & 1.26 & 1.84 & 1.73 & 6.47 & 1.38 & 1.86 & 2.95 & 2.10 \\
\midrule
\multicolumn{9}{c}{\textbf{Multimodal}} \\
\midrule
Method & Rele. & Infl. & Dive. & Uniq. & Clic. & \makecell{Sub.\\Posi.} & \makecell{Sub.\\Volu.} & Aver. \\
\midrule
Tran\_SEO  & -0.05 & 0.44 & 2.22 & 1.94 & 0.86 & 0.46 & 0.76 & 0.64 \\
Flue\_Expr & 1.03 & 1.53 & 1.83 & 6.03 & 1.64 & 1.51 & 2.43 & 1.85 \\
Quat\_Addi & 1.25 & 1.68 & 2.58 & 6.36 & 1.95 & 1.67 & 2.70 & 2.21 \\
Stat\_Addi & -0.41 & -0.77 & 1.92 & -1.06 & 1.42 & -0.91 & -2.41 & -0.58 \\
Capt\_Inje & 0.78 & 1.81 & 1.79 & 6.18 & 1.15 & 1.90 & 2.71 & 1.90 \\
\bottomrule
\end{tabularx}
\end{table}
\begin{table}[t]
\centering
\caption{Relative improvement (\%) on subjective visibility sub-dimensions of Caption Injection and text-only baselines on the MRAMG-Web dataset.}
\label{tab7}
\begin{tabularx}{\textwidth}{l *{8}{>{\centering\arraybackslash}X}}
\toprule
\multicolumn{9}{c}{\textbf{Unimodal}} \\
\midrule
Method & Rele. & Infl. & Dive. & Uniq. & Clic. & \makecell{Sub.\\Posi.} & \makecell{Sub.\\Volu.} & Aver. \\
\midrule
Tran\_SEO  & -2.48 & -2.21 & -1.43 & -6.39 & -3.12 & -3.10 & -2.62 & -3.05 \\
Flue\_Expr &  0.04 &  0.03 &  0.12 &  0.07 &  0.26 & -0.01 &  0.06 &  0.04 \\
Quat\_Addi &  0.28 &  0.35 &  0.67 &  0.80 &  0.46 &  0.17 &  0.91 &  0.46 \\
Stat\_Addi & -3.43 & -3.21 & -1.76 & -9.93 & -4.56 & -5.12 & -3.62 & -4.50 \\
Capt\_Inje & -0.31 & -0.16 & -0.49 &  0.16 & -0.78 &  0.03 & -0.51 & -0.36 \\
\midrule
\multicolumn{9}{c}{\textbf{Multimodal}} \\
\midrule
Method & Rele. & Infl. & Dive. & Uniq. & Clic. & \makecell{Sub.\\Posi.} & \makecell{Sub.\\Volu.} & Aver. \\
\midrule
Tran\_SEO  & -1.64 & -1.39 & -0.93 & -4.22 & -2.12 & -1.53 & -1.93 & -1.96 \\
Flue\_Expr & -0.14 & -0.09 &  0.02 & -0.05 &  0.08 &  0.01 & -0.07 & -0.07 \\
Quat\_Addi &  0.24 &  0.54 &  0.94 &  0.99 &  0.54 &  0.29 &  1.09 &  0.61 \\
Stat\_Addi & -2.29 & -2.33 & -1.44 & -7.13 & -2.73 & -2.81 & -3.17 & -3.10 \\
Capt\_Inje &  0.44 &  1.17 &  1.71 &  3.11 &  0.96 &  0.70 &  2.19 &  1.33 \\
\bottomrule
\end{tabularx}
\end{table}
\begin{table}[t]
\centering
\caption{Relative improvement (\%) on subjective visibility sub-dimensions of Caption Injection and text-only baselines on the MRAMG-Wiki dataset.}
\label{tab8}
\begin{tabularx}{\textwidth}{l *{8}{>{\centering\arraybackslash}X}}
\toprule
\multicolumn{9}{c}{\textbf{Unimodal}} \\
\midrule
Method & Rele. & Infl. & Dive. & Uniq. & Clic. & \makecell{Sub.\\Posi.} & \makecell{Sub.\\Volu.} & Aver. \\
\midrule
Tran\_SEO  & -0.30 & -0.15 &  1.14 & -1.23 &  0.38 & -0.52 &  0.10 & -0.14 \\
Flue\_Expr &  0.10 &  0.29 &  0.29 &  1.31 &  0.30 &  0.31 &  0.68 &  0.36 \\
Quat\_Addi &  0.47 &  0.86 &  0.89 &  2.63 &  0.87 &  0.60 &  1.31 &  0.97 \\
Stat\_Addi & -1.24 & -1.73 & -0.44 & -5.32 & -0.54 & -1.68 & -2.47 & -1.93 \\
Capt\_Inje &  0.33 &  0.52 & -0.15 &  3.11 &  0.39 &  0.36 &  1.25 &  0.70 \\
\midrule
\multicolumn{9}{c}{\textbf{Multimodal}} \\
\midrule
Method & Rele. & Infl. & Dive. & Uniq. & Clic. & \makecell{Sub.\\Posi.} & \makecell{Sub.\\Volu.} & Aver. \\
\midrule
Tran\_SEO  & -0.64 & -0.43 &  0.55 & -1.88 & -0.21 & -0.53 & -0.61 & -0.57 \\
Flue\_Expr &  0.12 &  0.45 &  0.66 &  1.72 &  0.48 &  0.30 &  0.95 &  0.58 \\
Quat\_Addi &  0.48 &  0.83 &  1.04 &  2.65 &  0.80 &  0.67 &  1.32 &  1.02 \\
Stat\_Addi & -1.42 & -1.72 & -0.31 & -4.29 & -0.91 & -1.40 & -2.23 & -1.77 \\
Capt\_Inje &  0.60 &  1.00 &  0.66 &  4.40 &  0.67 &  0.82 &  1.90 &  1.24 \\
\bottomrule
\end{tabularx}
\end{table}
\begin{table}[t]
\centering
\caption{Relative improvement (\%) on subjective visibility sub-dimensions of Caption Injection and text-only baselines on the MRAMG-Wit dataset.}
\label{tab9}
\begin{tabularx}{\textwidth}{l *{8}{>{\centering\arraybackslash}X}}
\toprule
\multicolumn{9}{c}{\textbf{Unimodal}} \\
\midrule
Method & Rele. & Infl. & Dive. & Uniq. & Clic. & \makecell{Sub.\\Posi.} & \makecell{Sub.\\Volu.} & Aver. \\
\midrule
Tran\_SEO  & -0.98 & -1.45 &  0.10 & -4.80 & -0.07 & -1.80 & -1.88 & -1.58 \\
Flue\_Expr & -0.13 &  0.11 &  0.30 &  0.78 &  0.27 &  0.01 &  0.06 &  0.09 \\
Quat\_Addi &  0.24 &  0.77 &  0.69 &  1.56 &  0.60 &  0.50 &  1.34 &  0.70 \\
Stat\_Addi & -1.65 & -3.53 & -2.05 & -9.63 & -1.00 & -3.40 & -4.70 & -3.63 \\
Capt\_Inje &  0.28 &  0.73 &  0.13 &  1.98 &  0.32 &  0.44 &  1.24 &  0.61 \\
\midrule
\multicolumn{9}{c}{\textbf{Multimodal}} \\
\midrule
Method & Rele. & Infl. & Dive. & Uniq. & Clic. & \makecell{Sub.\\Posi.} & \makecell{Sub.\\Volu.} & Aver. \\
\midrule
Tran\_SEO  & -0.72 & -0.71 &  1.25 & -2.15 &  0.21 & -0.95 & -1.05 & -0.65 \\
Flue\_Expr & -0.22 & -0.22 &  0.45 & -0.10 &  0.13 & -0.21 & -0.39 & -0.15 \\
Quat\_Addi &  0.24 &  0.52 &  1.17 &  1.21 &  0.83 &  0.38 &  0.67 &  0.65 \\
Stat\_Addi & -1.25 & -2.20 & -0.28 & -5.50 & -0.53 & -1.92 & -3.43 & -2.15 \\
Capt\_Inje &  0.65 &  1.86 &  1.20 &  3.88 &  0.88 &  0.99 &  3.04 &  1.60 \\
\bottomrule
\end{tabularx}
\end{table}
\clearpage
\section{Complete Experimental Results on MRAMG}
As shown in Tab.~\ref{tab4}--\ref{tab9}, we present the complete experimental results of Caption Injection and other text-only baseline methods across the MRAMG datasets, including the relative improvement scores across for all subjective visibility sub-dimensions.

\end{document}